\documentclass[11pt,twoside]{article}
\usepackage{asp2010}

\resetcounters

\newcommand{\teff}{\ensuremath{T_{\rm{eff}}}}
\newcommand{\pdot}{\ensuremath{\dot P}}
\newcommand{\uHz}{\ensuremath{\mu{\rm{Hz}}}}
\newcommand{\kep}{{\em Kepler}}
\newcommand{\kepmi}{{\em Kepler Mission}}
\newcommand{\mkep}{{\it Kp}}

\newcommand{\target}{KIC\,8626021}
\newcommand\rel[2]{{{#1}\,=\,{#2}}}
\newcommand\ellone{\rel{$\ell$}{1}}
\newcommand\elltwo{\rel{$\ell$}{2}}

\begin{document}

\title{Nine months of monitoring of a V777-Her pulsator\\
       with the Kepler spacecraft}
\author{Roy~H.~{\O}stensen}
\affil{Instituut voor Sterrenkunde, KU Leuven,
       B-3001 Leuven, Belgium}

\begin{abstract}
The V777 Her pulsator KIC 8626021 has been observed continuously by the Kepler space telescope for nine months.
Two new independent pulsation modes are detected in the extended dataset, which both match the predicted
sequence of \elltwo\ modes.
We demonstrate the photometric stability of the main pulsation modes and discuss the prospect of measuring
period changes within the context of the extended Kepler mission.
\end{abstract}

\section*{Introduction}

The \kep\ spacecraft is monitoring a 105 deg$^2$ field in the Cygnus--Lyrae
region, primarily to detect transiting planets \citep{borucki11}.
In the first four quarters of the \kepmi, a survey for pulsating stars
was made, and a total of 113 compact-pulsator
candidates were checked for variability \citep{ostensen10b,ostensen11b}.
The survey was extremely successful with respect to subdwarf-B (sdB) pulsators,
with discoveries including one clear V361-Hya pulsator \citep{kawaler10a},
a total of thirteen V1093-Her stars \citep{reed10a,kawaler10b,baran11b},
including a spectacular sdB+dM eclipsing binary in which the hot primary
shows an exceptionally rich pulsation spectrum \citep{2m1938}.
However, not a single pulsating white dwarf was found during the survey
phase. As the \kep\ Guest Observer programme permits monitoring of
interesting targets in one month or longer slots, we started searching among
the faint rejects from our original survey sample, where the \mkep\,=\,18.46\,mag
\target\ was found. The first 1-month dataset of this V777-Her pulsator
was obtained in the second month of quarter 7 of the \kepmi\ (Q7.2),
and the Fourier transform (FT)
revealed eleven peaks, of which nine forms triplets with even splittings of
3.3\,\uHz, corresponding to a rotation period of 1.7\,d when assuming
that the modes have degree \ellone. For further details about the search
and the Q7.2 dataset, please see the discovery paper; \citet{ostensen11c}.

An asteroseismic solution for \target\ was presented by \citet{bischoff-kim11},
where all the modes could be well matched by a structural model only if
the target is significantly hotter than the spectroscopic estimate made in the
discovery paper. Since the spectrum had a rather low signal-to-noise ratio, and
DB stars are notoriously difficult to get reliable temperatures for,
a higher temperature is not controversial. A further asteroseismic study
by \citet{corsico12} corroborates the high temperature solution.

\begin{table}[t]
\caption{Frequencies detected in the Q10+11+12 dataset.}\label{tbl:freq}
\smallskip
\begin{center}\small
\begin{tabular}{lllll}\tableline\noalign{\smallskip}
ID & Frequency & Period & Type & $\ell,k$ \\
\noalign{\smallskip}\tableline\noalign{\smallskip}
$f_1$ & 4309.89 & 232.02 & Triplet, $\delta f=3.3\,\mu$Hz & 1,4 \\
$f_2$ & 5070.05 & 197.11 & Triplet, $\delta f=3.3\,\mu$Hz & 1,3 \\
$f_3$ & 3681.87 & 271.60 & Triplet, $\delta f=3.3\,\mu$Hz & 1,5 \\
$f_4$ & 3294.22 & 305.56 & Single peak & 1,6 \\
$f_5$ & 2658.85 & 376.10 & Single peak & 1,8 or 2,15 \\
$f_6$ & 6965.29 & 143.57 & Single peak & 2,4 \\
$f_7$ & 4398.37 & 227.36 & Single peak & 2,8 \\
\noalign{\smallskip}\tableline
\end{tabular} \end{center} \small
\end{table}

\section*{The 9-month dataset}

\begin{figure*}[!ht]
\centering
\includegraphics[width=13.0cm,angle=0]{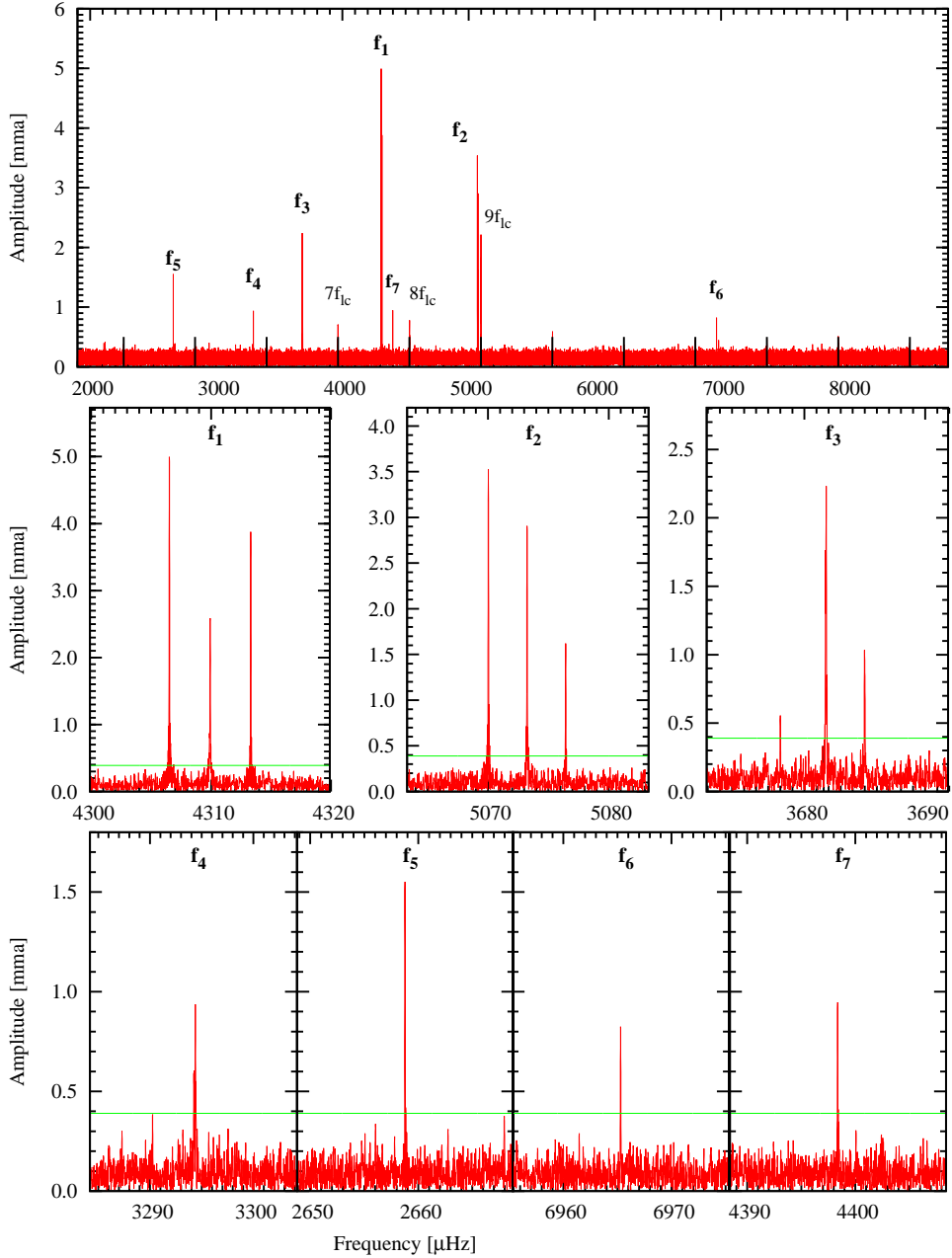}
\caption{
Fourier spectrum of \target\ based on the full Q10+11+12 dataset.
The 7 independent modes are marked as $f_1$ to $f_7$, see Table~\ref{tbl:freq}.
Artifacts
produced by the long cadence cycle ($f_{\rm lc}$\,=\,566.423\,\uHz) are also marked.
}
\label{fig:ft}
\end{figure*}

\kep\ quarters Q10, Q11 and Q12 covers BJD 2455739.83 to 2456015.03, giving a full
span of 275.2 days, which corresponds to a frequency resolution of $\sim$0.04\,\uHz.
The dataset contains 389,195 60-s measurements, or 386,863 after sigma clipping.
The eight monthly data-downlink gaps are 0.74 to 1.06\,d in duration, and 0.87\,d on average,
which is too short to produce significant monthly aliases in the Fourier spectrum.
Since the target is so faint, only $\sim$250\,e$^-$/s are collected,
corresponding to $\sim$15,000 counts per short cadence measurement. In signal-to-noise terms
we get only S/N\,=\,30 (Q10), 32 (Q11), and 28 (Q12). Still, with 43,000 consecutive
measurements per month, this is sufficient to bring down the mean Fourier amplitude
level to $\sim$0.25\,mma in a month of data, and to just below 0.1\,mma in the 9-month dataset
(Fig.~\ref{fig:ft}).

Running Fourier spectra were made by clipping the full Q10+11+12 dataset
into chunks of up to 14\,d length and taking the FT of each. When displayed as
a grey-scale plot as in Fig.~\ref{fig:running} it is easy to see that some of
the peaks such as the central component of the $f_1$ triplet variable amplitudes.
Other peaks, such as the longest-frequency component of the $f_1$ and $f_2$
triplets, show no significant amplitude variability over the course of the
nine-month run.

\section*{New frequencies}

\begin{figure*}[!h]
\centering
\includegraphics[height=13.5cm,angle=-90]{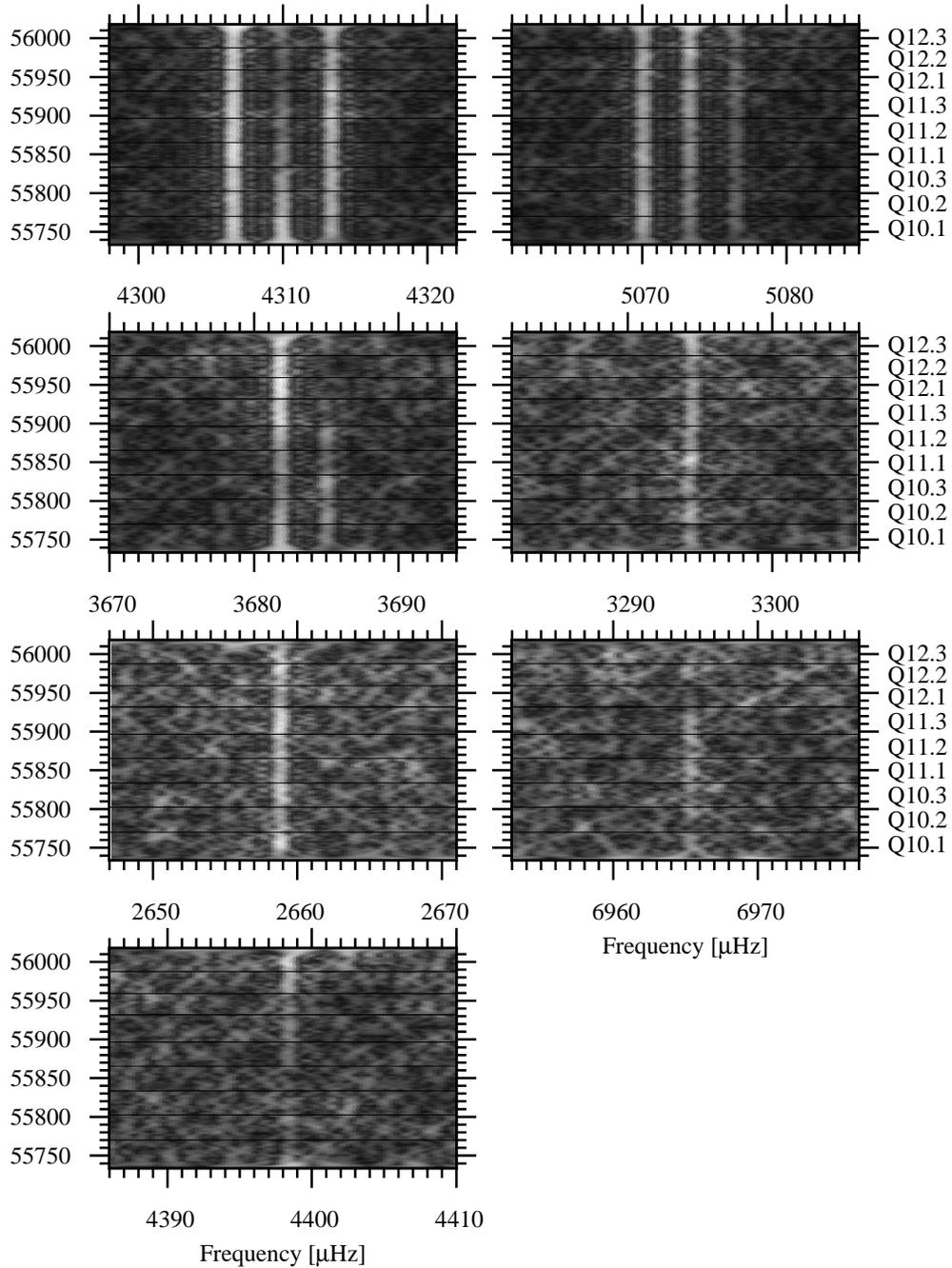}
\caption{
Running Fourier spectra of \target\ produced by clipping the full Q10+11+12 light curve
into chunks of up to 14\,d length and taking the FT of each.
Time in BJD\,+\,2400000 is given on the left-hand axis and \kepmi\ months are indicated
on the right-hand side. Horizontal lines mark the monthly Earth downlink interruptions.
}
\label{fig:running}
\end{figure*}

In addition to the five independent modes seen in Q7.2, two new ones are detected in the 9-month dataset,
that were below the 4-$\sigma$ detection limit in the discovery run.
These are labeled $f_6$ and $f_7$ in Table~\ref{tbl:freq}
and Figures~\ref{fig:ft} and \ref{fig:running}.
On closer inspection, $f_6$ is present in Q7.2 with a slightly higher amplitude than in the 9-month
dataset, but 0.85\,mma is only 3.5\,$\sigma$ in Q7.2, well below our preferred detection
limit of 4\,$\sigma$. $f_7$ is below the 3-$\sigma$ level in Q7.2.
In the 9-month FT, $f_6$ has S/N\,=\,8.13 and $f_7$ has S/N\,=\,8.48. 
That the two new frequencies are real and stable pulsation modes of the star is immediately clear from
the running FT. While well below the detection limit in individual FTs, their steady
recurrence at the same frequency betrays their presence.

The spacing for \ellone\ modes derived in the discovery paper was $\Delta P_1$\,=\,35.7\,s. 
Since $\Delta P_1$/$\Delta P_2$ = $\sqrt{\scriptstyle 3}$ for $g$-modes in the asymptotic limit,
one can assume an \elltwo\ spacing of $\Delta P_2$\,=\,20.6\,s.
In the discovery paper $f_5$ was identified as \ellone, $k$\,=\,8, whereas in
\citet{bischoff-kim11} it was found to be a closer match to \elltwo, $k$\,=\,15.
Assuming the latter, one can tentatively assign $f_6$ as \elltwo, $k$\,=\,4, and
$f_7$ as \elltwo, $k$\,=\,8.
This is very encouraging as the number of free parameters in the asteroseismic models had to be 
kept small with only the original five independent modes to work from. 
Adding two more independent modes, allows the models to be substantially improved.

\section*{Conclusions}

A faint V777-Her pulsator was found in the \kep\ field, and 
observed for one month during Q7, and monitored continuously
from Q10 onwards.
The ongoing observations have revealed that some of the strongest modes
are almost perfectly stable in time, and that low-amplitude modes
come and go.
Further monitoring should therefore reveal more independent modes, thereby
providing further constraints on the asteroseismic models.

With the current mission extension, we have the opportunity to get
more than five years of near-continuous monitoring, provided the spacecraft
maintains its excellent performance.
With such a time-base, we have a unique opportunity to detect
period changes (\pdot) due to the changes in interior structure as the white dwarf cools.

The rate of period change in a white dwarf has so far only been measured
in the ZZ-Ceti pulsator, RY\,LMi, for which \citet{kepler05}
found \pdot\,=\,(3.57$\pm$0.82)\,$\times$\,10$^{-15}$\,s/s based on
31 years of monitoring. At the high-\teff\ end of the V777-Her instability strip
the cooling is completely dominated by plasmon neutrino emission \citep{winget04},
and is predicted to be a hundred times faster than for the DAVs
due to the neutrino contribution.
The measurement of an evolutionary \pdot\ is therefore quite possible within the
time-frame of the extended {\em Kepler Mission}, as currently scheduled to go
on until the end of 2016.

\acknowledgements
The research leading to these results has received funding from the European
Research Council under the European Community's Seventh Framework Programme
(FP7/2007--2013)/ERC grant agreement N$^{\underline{\mathrm o}}$\,227224
({\sc prosperity}), as well as from the Research Council of K.U.Leuven grant
agreement GOA/2008/04.

The author gratefully acknowledges the \kep\ team and everybody who
has contributed to making the mission possible.
Funding for the \kepmi\ is provided by NASA's Science Mission Directorate.

\bibliography{sdbrefs}

\end{document}